\documentclass{iopconfser}

\usepackage{a4wide,amssymb,cite,graphicx}
\usepackage{epsfig}
\usepackage{hyperref}

\usepackage{wrapfig}   % Allows floating figures on left/right
\usepackage{bm}

\begin{document}

\noindent Conference Proceedings for BCVSPIN 2024: Particle Physics and Cosmology in the Himalayas\\Kathmandu, Nepal, December 9-13, 2024 

\title{What is AI, what is it not,  how we use it in physics and how it impacts... you}

\author{Claire David$^{1}$}

\affil{$^1$African Institute for Mathematical Sciences (AIMS),  Cape Town,  South Africa}

\email{claired@aims.ac.za}

\begin{abstract}  
\\[1ex] Artificial Intelligence (AI) and Machine Learning (ML) have been prevalent in particle physics for over three decades, shaping many aspects of High Energy Physics (HEP) analyses. As AI's influence grows, it is essential for physicists --- as both researchers and informed citizens --- to critically examine its foundations, misconceptions, and impact. This paper explores AI definitions, examines how ML differs from traditional programming, and provides a brief review of AI/ML applications in HEP, highlighting promising trends such as Simulation-Based Inference, uncertainty-aware machine learning, and Fast ML for anomaly detection. Beyond physics, it also addresses the broader societal harms of AI systems, underscoring the need for responsible engagement. Finally, it stresses the importance of adapting research practices to an evolving AI landscape, ensuring that physicists not only benefit from the latest tools but also remain at the forefront of innovation.
\end{abstract}

\newpage

\section{Introduction}
Machine learning (ML) techniques have been used in particle physics for the past three decades.  However, with the technological leap that occurred during the past decade in generative Artificial Intelligence (AI),  there is now a growing need to ask ourselves critical relevant questions and take a stance on present and future AI deployments, both within and outside the field of particle physics. 

After exploring the definitions of the key terms of AI and explaining the drastic conceptual changes brought by AI in Section~2 (``What is AI?"),  Section~3 will consider the common misconceptions about the field (``AI: What it is not").  A succinct review of the usage of AI and ML in particle physics is presented in the section 4. Section 5 steps outside the realm of physics to examine the individual and societal impacts of AI tools in our daily lives. The document concludes with some reflections on how we might navigate the ongoing AI revolution.

\section{What is AI?}
What does the somewhat provocative term `artificial intelligence' mean? What is it and what is its scope? Is the terminology borrowed from human behaviour truly appropriate? This section challenges the reader with deceptively simple questions, presents the paradigm shift AI has introduced through the data-driven approach of machine learning, and demystifies the core mathematics behind most AI algorithms.

\subsection{A few basic questions}
Defining the key terms of artificial intelligence is challenging because it presupposes that there is agreement about the meaning of intelligence.  \textit{What is intelligence?}  Intelligence is a complex and widely debated concept, with a pluralistic nature involving abstraction, logic, emotional processing, memory, and more. Its definition varies across disciplines, from psychology to neuroscience and philosophy, depending on the perspective taken.  In~\cite{legg2007universal},  Shane Legg and Marcus Hutter attempt to provide a universal definition of intelligence. However, the very first sentence of their article reads: \\
\hspace{3ex}``\textit{A fundamental problem in artificial intelligence is that nobody really knows what intelligence is.}"\\
They eventually settle with an informal working definition:\\
\hspace{3ex}\textit{``Intelligence measures an agent's ability to achieve goals in a wide range of environments."}\\
One common critique of this definition is that ``a system may appear to be intelligent without really understanding anything"~\cite{searle1980minds}. 

This leads us to ask: \textit{``What is understanding?”} And, as it appears in machine learning, we must also ask: \textit{``What is learning, and how do the two differ?”} Understanding implies deeper insights and a stronger grasp of knowledge, whereas learning is about the acquisition of knowledge. The key takeaway here is to question whether it is legitimate to borrow terms that are rooted in human behaviour when these terms themselves lack clear and widely accepted definitions.

\subsection{The paradigm shift from explicit programming}
\begin{figure}[!h]
\centering
\includegraphics[width=0.8\linewidth]{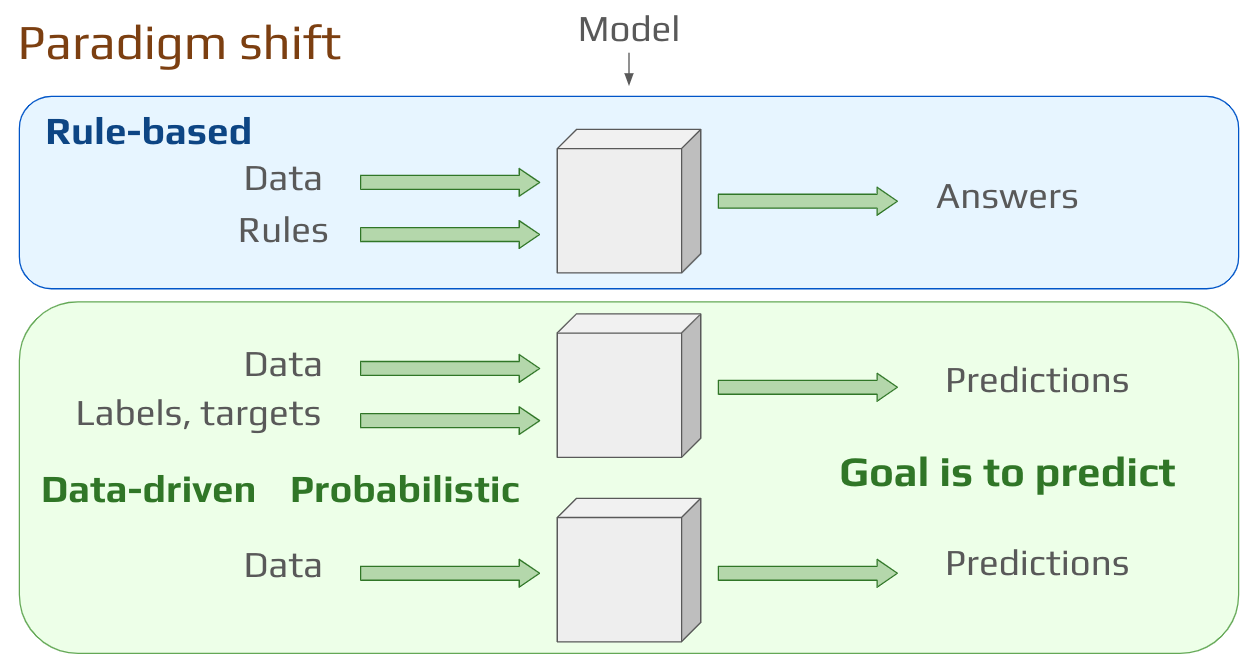} 
\caption{The fundamental change from explicit programming (top in blue) versus machine learning (upper green,   supervised machine learning, lower green, unsupervised machine learning).  Illustration from the author.}
\label{fig:paradigm_shift}
\end{figure}

What machine learning brings to the computer world is a drastic paradigm shift. Before the advent of ML algorithms,  programming referred to sets of instructions prepared by a human programmer.  These could be dynamic and complex with different outcomes depending on inputs fulfilling or not conditional statements (\texttt{if},  \texttt{while},  \texttt{for} etc). Yet these outcomes are limited by design. Explicit programming is ruled-based.  By contrast,  supervised machine learning uses the `answers' (that is,  the targets) and other data, to derive implicit rules from the data in order to make predictions (a proxy for answer).  In the context of unsupervised learning, implicit rules are derived from data without the aid of targets (see Figure~\ref{fig:paradigm_shift}).  

Such a paradigm shift brings the focus to the data --- a fortunate turn for physics, as data ultimately holds the answers.

\subsection{The core mathematics of AI/ML}
As complex and sophisticated as modern AI can be,  the mathematics at the core of the majority of those algorithms are simple: it's all about fitting a function to the data.  Deep down, it is a minimization problem.  Mathematically, given a dataset of features $\bm{x}$ and targets $\bm{y}$,  
\begin{equation}
D = {(x_i, \: y_i)}, i = 1,   \cdots  n  \;, 
\end{equation}
fitting the data --- that is training --- is choosing the hypothesised function $h$* that minimises the empirical risk $\hat{R}$ --- that is averaged loss ${L}$,  or cost --- defined as: 
\begin{equation}
\hat{R}(h) = \frac{1}{n} \sum_i^n L(h(x_i),\,y_i)  \;.
\label{eq:minrisk}
\end{equation}
The hypothesised function needs to be sufficiently flexible (which is usually the case with a deep neural network).  It is important to note that the optimal $h$* only depends on the form of the loss function and the distribution of the data\cite{Prosper:2024wjh}.  
The most advanced algorithms may involve functions of immense complexity, yet at their core, the underlying mathematical operations are governed by the general Equation~\ref{eq:minrisk}. 

\section{AI: What it is not}
The field of AI is interdisciplinary, built on mathematics, statistics, and computer science, while also drawing concepts from physics.  
However, it is not ``glorified statistics" as some may argue on the internet. The goals differ: statistics are geared more toward an ``autopsy" of the data,  studying correlations and making inferences, whereas AI  is largely about predictions.   

Most AI deployed as tools for physics do not attempt physics modelling.  For a theory in physics to be complete, it must be both descriptive and predictive.  Machine learning accomplishes the latter but does not achieve the former at present.  AI is not ``the solution"  for our research but rather a tool for a given, precise task.  
However, with the advent of Large Language Models (LLMs), researchers at all levels now use them for tasks like writing papers, coding, or even as self-teaching tools. But caution is warranted because these agents are still prone to hallucinations, a colloquialism for nonsensical outputs.

\section{AI in Physics}

\subsection{Brief overview of AI in HEP}
\begin{wrapfigure}{r}{0.5\textwidth} % 'r' for right-side placement, adjust width as needed
    \vspace{-15pt}
    \centering
    \includegraphics[width=\linewidth, keepaspectratio]{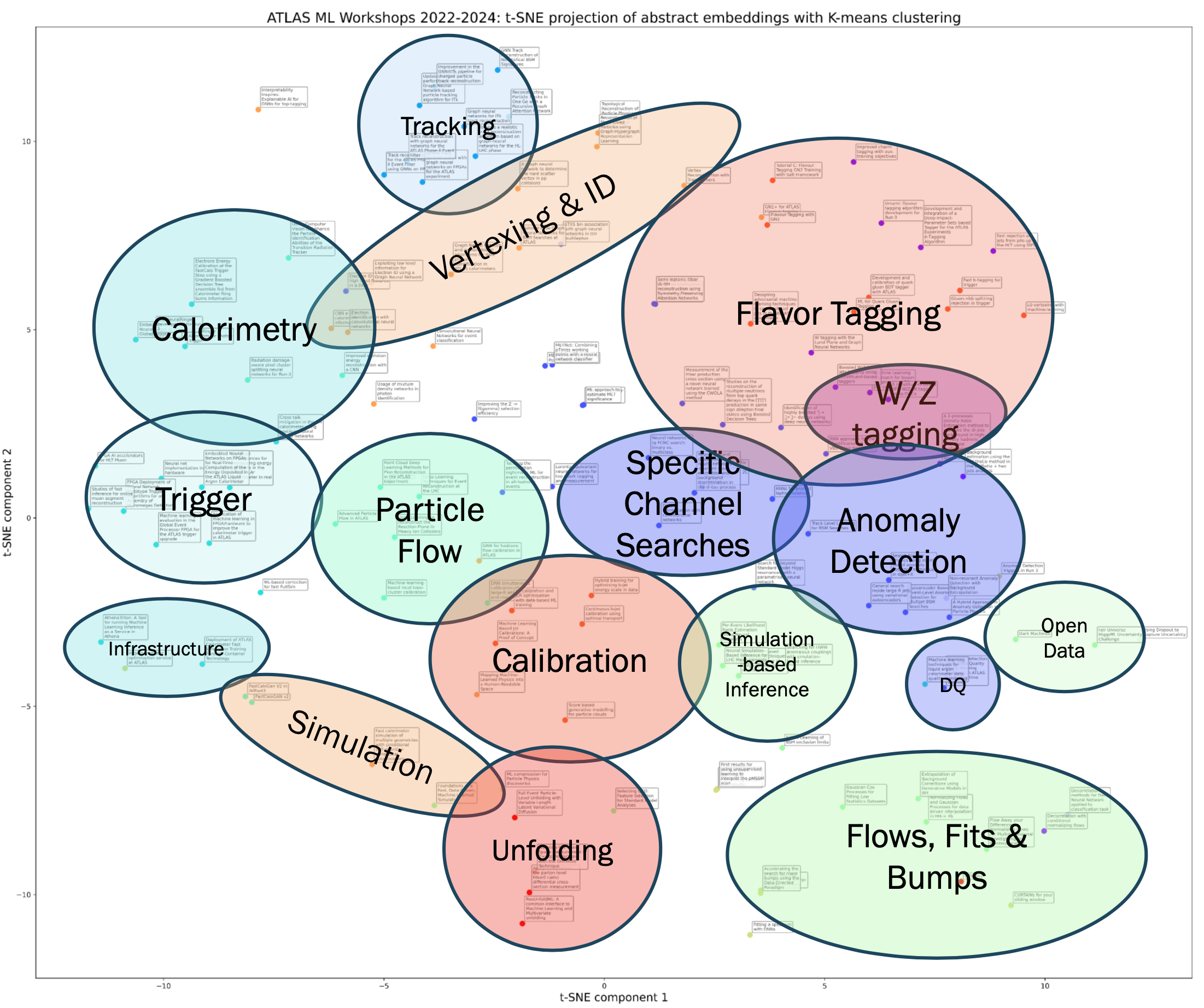}
    \caption{t-SNE projection of abstract embeddings with K-means clustering for ATLAS Workshop notes between 2022 and 2024~\cite{talk_murnane_ml_hep}.}
    \label{fig:davidc_ml_hep_murnane}
    \vspace{15pt}
        \centering
    \includegraphics[width=0.5\textwidth, keepaspectratio]{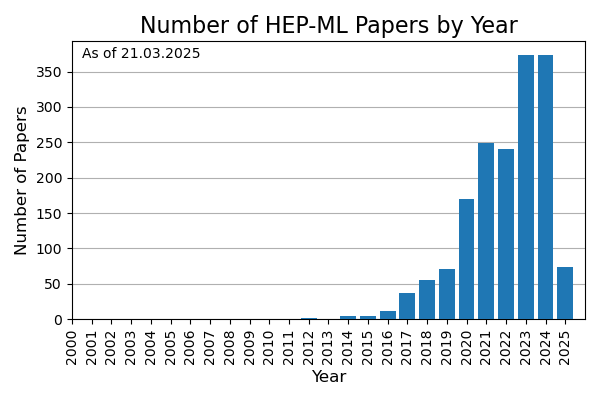}
    \caption{Distribution of the papers in High Energy Physics featuring deep learning approaches to experimental, phenomenological, or theoretical analyses.  Source: HEP ML Living Review~\cite{hepmllivingreview}.}
    \label{fig:davidc_ml_paper_per_year}
    \vspace{-15pt}
\end{wrapfigure}

AI/ML is now ubiquitous in High Energy Physics (HEP) and the applications are wide-ranging and far too numerous to review here. Instead we highlight a few key trends.  Figure~\ref{fig:davidc_ml_hep_murnane} provides an overview focused solely on the ATLAS Experiment and only covers the past few years: ML techniques are deployed throughout each step of an analysis workflow.  From the initial stages of triggering, vertexing, and track fitting to object tagging, unfolding, signal extraction and final parameter inference. AI is everywhere.

With the number of HEP-ML paper exponentially growing (see Figure~\ref{fig:davidc_ml_paper_per_year}), it is difficult to keep track of developments. Happily,  the ``HEP ML Living Review"\cite{hepmllivingreview} is an online collection of publications for modern machine learning applied to particle physics,  grouped into categories for easier browsing.  A similar initiative exists in cosmology under the GitHub repository \texttt{ml-in-cosmology}\cite{mlincosmology,george_stein_2020_4024768}.  

%\begin{wrapfigure}{l}{0.5\textwidth} % 'r' for right-side placement, adjust width as needed
%    \vspace{-15pt}
%    \centering
%    \includegraphics[width=0.5\textwidth, keepaspectratio]{davidc_ml_paper_per_year.png}
%    \caption{Distribution of the papers in High Energy Physics featuring deep learning approaches to experimental, phenomenological, or theoretical analyses.  Source: HEP ML Living Review~\cite{hepmllivingreview}.}
%    \label{fig:davidc_ml_paper_per_year}
%    \vspace{-10pt}
%\end{wrapfigure}

\subsection{The latest trends}
In recent years, machine learning for HEP has progressed in promising directions aimed at addressing the main challenges in the field. 

Simulation-Based Inference (SBI) offers a statistical approach that does not require explicit knowledge of the likelihood function. There is no longer a need to reduce high-dimensional data to summary statistics. SBI allows for direct analysis of high-dimensional data using simulated data and machine learning~\cite{Brehmer:2019xox}.

Another recent and promising avenue is uncertainty-aware machine learning.  Instead of training a classifier first and then assessing the systematic uncertainty independently,  the new method uses a profiled-classifier explicitly dependent on nuisance parameters (by construction).  The dependence on the uncertainties actually increases the sensitivity of the analysis to parameters of interest~\cite{Ghosh_2021}. In October 2024, the ``FAIR Universe – NeurIPS 2024 Higgs Uncertainty Challenge" was launched on Codabench~\cite{FAIRUniverse,FAIR_Unc_Challenge_2024}.  In this competition,  participants are invited to design an advanced analysis technique that can not only measure the signal strength but also provide a confidence interval (the contest closed on March 14, 2025 and the results were unavailable at the time of this writing).

In the absence of strong, preferred theoretical directions, so-called model-independent approaches have become fashionable.  Anomaly detection, which is not a specific algorithm but rather a task, has become increasingly popular because of its signal-agnostic nature. It can be implemented using various methods, including decision trees, generative adversarial networks (GANs), and normalizing flows. However, the most commonly used techniques are Autoencoders (AEs) and Variational Autoencoders (VAEs), which excel at learning representations of standard physics and flagging departures from it.  With the recent progress in accelerator hardware,  a growing community of physicists, engineers, and computer scientists are developing real-time, low-latency, on-detector ML algorithms, usually deployed for triggering purposes~\cite{fastml}.

\subsection{A few questions to physicists and humans}
Do we want more black boxes? Are we overcomplicating analyses? In this era of AI prevalence,  jumping into  AI tools is tempting.  While AI surely augment the performance of both experimental and theoretical physics, it comes at a cost. The research driven by AI inherently sacrifices \textit{interpretability}.  An interpretable model can be understood by a human without any other aids --- that is the ``how" predictions are made.  But most AI algorithms deployed in physics, such as deep neural networks,  random forests, gradient boosting machines,  are not interpretable in the sense that their implicit rules cannot be made intelligible.  Yet,  they can still be explainable using post-hoc techniques. \textit{Explainability} relates to the ``why" an algorithm made a particular prediction.  

It could be unsettling for scientists to introduce extra steps into their workflow just to explain an answer we cannot interpret. Are we giving up on the physics? Does a neural network truly grasp the core meaning of our kinematic variables? A broader introspection on this challenge could open new avenues, such as embedding prior physics knowledge into AI tools to harness the best of both worlds: organic intuition from human physicists and advanced data-driven insights from AI.

\section{AI and You}
Beyond the laboratory, AI has permeated our daily lives, making it impossible to remain unaffected by it.

We make no attempt to summarize here the full spectrum of both the great benefits and significant harms that AI brings.  Such a list would need to be constantly updated to reflect the latest innovations in this fast-evolving field.  But we can make some general remarks.

For any new algorithm or applications, it is crucial to ask ourselves questions: does this AI tool educate or manipulate? Does it help or does it overly assist? Does it harm, or even kill? These questions are essential to provide a guiding framework for receiving and analysing the incoming AI buzz. 

\begin{figure}[!h]
\includegraphics[width=\linewidth]{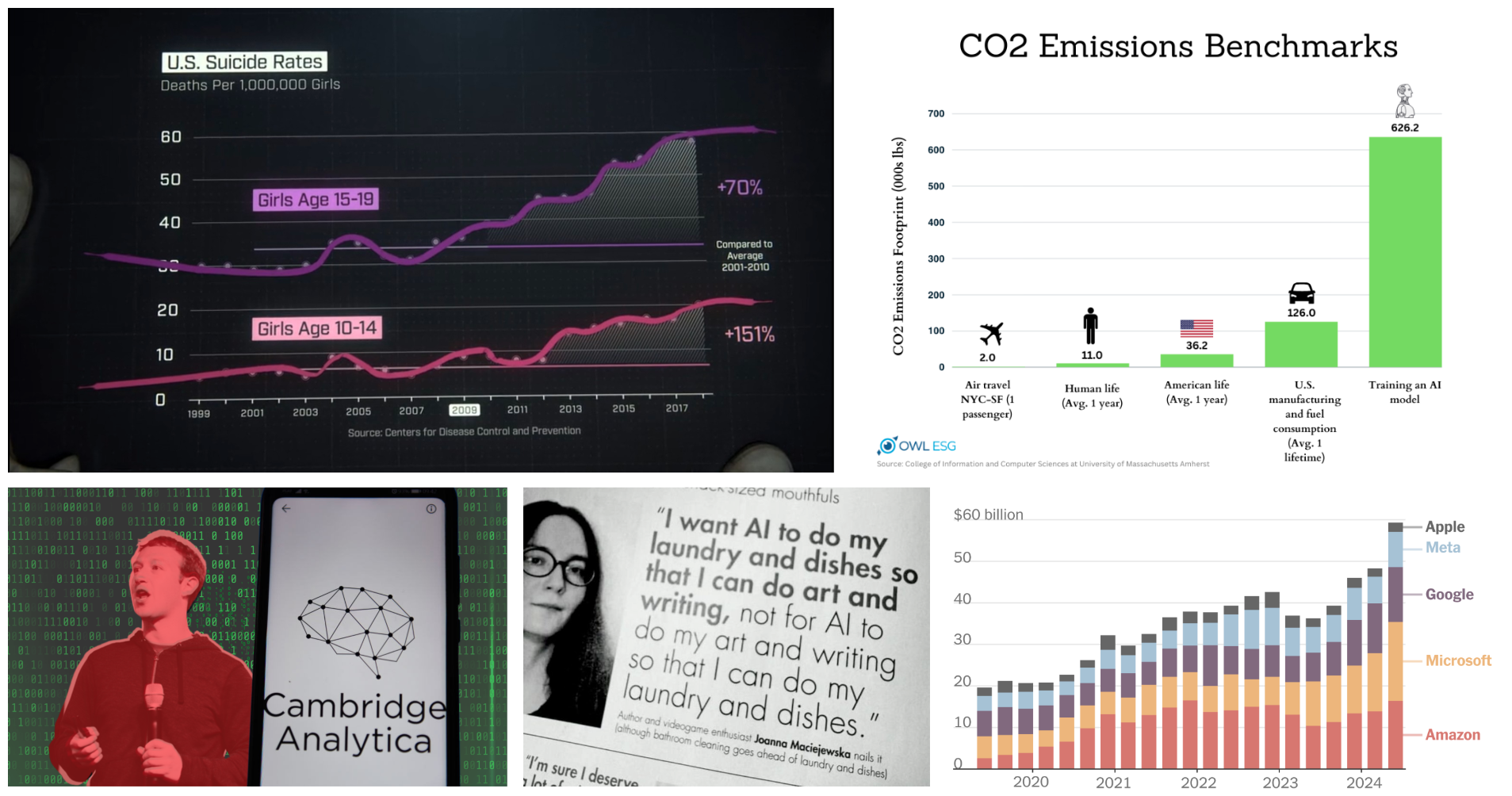} 
\caption{Illustrations highlighting various negative impacts of AI in society.  \\
Top left: \textit{The Social Dilemma},  Netflix, 2020.  Top right: OWL ESG \& College of Information and Computer Sciences,  University of Massachusetts Amherst~\cite{owlesg2024}. Bottom left: Cambridge Analytica scandal from www.performancemarketingworld.com~\cite{webarticle_cambridge_analytica}. Bottom middle: A viral quote from Joanna Maciejewska~\cite{maciejewska2024}.  Bottom right: FactSet and company filings,  Karen Weise,  The New York Times~\cite{NYTimes2024}.}
\label{fig:davidc_impact_ai}
\end{figure}

People have started to be vocal on the negative impacts of AI.  Some of these impacts --- far from all --- are illustrated on Figure~\ref{fig:davidc_impact_ai}.  Mainstream social media platforms use AI algorithms that are manipulative by design (the so-called attention market to keep users scrolling).  They exploit the vulnerability of users before exposing them to advertisements.  The docudrama ``The Social Dilemma" shows a graph of the suicide rates in the US among teenage girls; the soaring trends starts after the time social media became accessible on mobiles.  AI algorithms were also employed to construct psychological profiles and deliver targeted political ads. The Cambridge Analytica scandal~\cite{webarticle_cambridge_analytica} in 2018 revealed that the firm improperly harvested data from Facebook and used it specifically during the Brexit and 2016 U.S. presidential election campaigns, likely playing a significant role in shaping voter behaviour and influencing the final outcomes.  Another harm of AI that is quite understated is on the environment.  Generative AI in particular is incredibly polluting.  The training of such AI models release almost 60 times more CO$_2$ than the emission done by a human being on average in the course of an entire life~\cite{owlesg2024}.  Not to mention that the growing AI hype puts an increasing strain on the extraction of rare minerals, which has both environmental and ethical impacts.  Generative AI challenges us to rethink the essence of art and creation, as the quote from Joanna Maciejewska~\cite{maciejewska2024} aptly captures it. What should we use it for? Lastly,  big technology companies have increased their expenditures on AI,  approaching a total of 60 billion USD in 2024. For comparison, the gross domestic product (GDP) of Nepal stands at approximately 40 billion USD.

\subsection{Ground rules}
The AI revolution calls for a rethinking of the way physics is conducted.  This rethinking should take place at multiple levels, including institutions, research groups, and individual scientists.

Institutions should revise their undergraduate and graduate programs to incorporate lectures and training on the newest methods.  They should also allocate more computing resources, encourage interdisciplinary collaborations, and provide easy access to datasets.

Research groups should include in their work AI explainability studies such as LIME (Local Interpretable Model-agnostic Explanations) or SHAP (SHapley Additive exPlanations) or feature importance, which eventually serve in improving the interpretability of their research that uses AI.  In the context of experimental particle physics, analysis teams should justify the need of the extra complexity (and likely opacity) for each given AI deployment, and ideally compare with an interpretable aka ``white-box" model.  Also, the performance metrics commonly associated with AI techniques may not be relevant and meaningful in the context of physics: developing new metrics is thus essential. 

At the individual level, of course, it may be stating the obvious that studying physics and building a strong foundation is essential. To this day, human expertise and intuition remain unique and invaluable. AI is not yet competing with organic brain matter.  Another individual responsibility is to stay informed about the latest trends. This is quite a challenge in the rapidly expanding AI field.  However, platforms such as arXiv or Paper With Code offer easy access to recent developments and new algorithms.  Last but not least: Large Language Models have proved to be fantastic assistants helping academics with both scientific writing and programming. However, they can be a double-edged sword.  There is a risk of over-reliance and excessive delegation to AI tools, leading to missed opportunities to sharpen critical thinking, develop autonomy in reasoning, and master the fundamental skills needed to write independently or code proficiently. 

In this evolving landscape of numerous tempting tools and methods,  there need to be conversations across and within institutes and research groups,  ultimately leading to new policies on AI usage.  At the individual level, the question we must ask is: \textit{What are my ground rules regarding my usage of AI?} For instance,  it could be forbidding oneself from uploading documents or slides to an LLM and asking it to generate summaries or proceedings.  However, one might allow using such an assistant to rephrase selected sentences (as the author of this paper has done).  In this way, autonomous thinking prevails, and the practice of scientific writing remains in the hands of the human, not the AI agent.

\section{Takeaways}
Widely used in HEP for more than three decades,  AI techniques have helped optimize event selection and improve object identification, thus increasing the analysis sensitivity and enhancing physics discovery potential.  
%If deploying more machine learning has been met with reluctance or even criticism in the past,  
%If some deployments of AI were met with criticism or reluctance mostly due to an ever-increasing opacity in the analysis workflow,  the newest trends such as Simulation-Based Inference or uncertainty-aware machine learning are promising for our field.  
There has been reluctance in the community regarding the over-deployment of AI techniques, as they reduce the transparency of the analysis workflow.  However, the newest trends --- such as Simulation-Based Inference and uncertainty-aware machine learning --- offer promising avenues,  very well suited for the high-dimensional data characteristic of HEP. 

But AI is not doing the physics for us! At least not yet. With AI redefining the field,  there is an urgent need to adapt analysis pipelines and offer dedicated training to both early-career and senior researchers to keep up to date.  It is and will be a challenge --- albeit necessary --- to keep up-to-date with the latest AI advances. This requires additional time in the already fast-paced discipline that is HEP.  Collaborative efforts, such as the Inter-Experimental Machine-Learning (IML) Working Group from CERN or the Fast ML Lab collective, to name a few, are key in providing resources for physicists and fostering knowledge transfer. 

Last but not least, physicists utilizing AI in their research represent AI-informed citizens who ought to speak up about the  implications of AI both in and outside of physics.  Explaining what AI is, what it is not, how it is used in physics and how it impacts... everyone.

\section*{Acknowledgement}
The author would like to thank the international and local organizing committees for the invitation to give a talk at BCVSPIN 2024.  The author is also grateful to the Arthur B. McDonald Canadian Astroparticle Physics Research Institute for sponsoring her attendance to BCVSPIN 2024.

\bibliographystyle{jhep}
\bibliography{iopconfser_BCVSPIN2024_Claire_DAVID}{}

\providecommand{\href}[2]{#2}\begingroup\raggedright\begin{thebibliography}{10}

\bibitem{legg2007universal}
{Legg, S. and Hutter, M.}, \emph{{Universal Intelligence: A Definition of
  Machine Intelligence}},
  \href{http://dx.doi.org/10.1007/s11023-007-9079-x}{\emph{{Minds \& Machines}}
  {\bfseries 17} (2007) 391--444}.

\bibitem{searle1980minds}
{Searle, J.}, \emph{{Minds, Brains, and Programs}}, {\emph{{Behavioral \& Brain
  Sciences}} {\bfseries 3} (1980) 417--458}.

\bibitem{Prosper:2024wjh}
{Prosper, Harrison B.}, \emph{{Statistics and machine learning for high-energy
  physics}}, \href{http://dx.doi.org/10.23730/CYRSP-2025-002.151}{\emph{CERN
  Yellow Rep. School Proc.} {\bfseries 2} (2025) 151--195}.

\bibitem{talk_murnane_ml_hep}
D.~Murnane, ``{Overview of ML in ATLAS}.''
  {\url{https://indico.cern.ch/event/1445641/contributions/6092553/}},
  presented at {Machine Learning for Fundamental Physics School 2024}, Lawrence
  Berkeley National Laboratory, USA, 2024.

\bibitem{hepmllivingreview}
{{HEP ML Community}}, ``{A Living Review of Machine Learning for Particle
  Physics}.'' {\url{https://iml-wg.github.io/HEPML-LivingReview/}}, {Accessed:
  2025-03-31}.

\bibitem{mlincosmology}
{George Stein}, ``{Machine Learning in Cosmology}.''
  {\url{https://github.com/georgestein/ml-in-cosmology}}, {Accessed:
  2025-03-31}, 2020.

\bibitem{george_stein_2020_4024768}
{George Stein}, ``{georgestein/ml-in-cosmology: Machine learning in
  cosmology}.'' {\url{https://doi.org/10.5281/zenodo.4024768}}, {Accessed:
  2025-03-31}, 2020.
\newblock 10.5281/zenodo.4024768.

\bibitem{Brehmer:2019xox}
J.~Brehmer, F.~Kling, I.~Espejo and K.~Cranmer, \emph{{MadMiner: Machine
  learning-based inference for particle physics}},
  \href{http://dx.doi.org/10.1007/s41781-020-0035-2}{\emph{Comput. Softw. Big
  Sci.} {\bfseries 4} (2020) 3},
  [\href{https://arxiv.org/abs/1907.10621}{{\ttfamily 1907.10621}}].

\bibitem{Ghosh_2021}
A.~Ghosh, B.~Nachman and D.~Whiteson, \emph{Uncertainty-aware machine learning
  for high energy physics},
  \href{http://dx.doi.org/10.1103/physrevd.104.056026}{\emph{Physical Review D}
  {\bfseries 104} (Sept., 2021) }.

\bibitem{FAIRUniverse}
{FAIR Universe Collaboration}, ``{FAIR Universe Project}.''
  {\url{https://fair-universe.lbl.gov}}, Accessed: 2025-03-31, 2025.

\bibitem{FAIR_Unc_Challenge_2024}
{Codabench Competition \#2977 }, ``{FAIR Universe - HiggsML Uncertainty
  Challenge}.'' {\url{https://www.codabench.org/competitions/2977/}}, Accessed:
  2025-03-31, 2025.

\bibitem{fastml}
F.~M.~L. Community, ``Fast ml lab.'' {\url{https://fastmachinelearning.org/}},
  Accessed: 2025-03-31, 2025.

\bibitem{webarticle_cambridge_analytica}
{Performance Marketing World}, ``{Meta forced to pay \$725M in Cambridge
  Analytica data scandal}.''
  \url{https://www.performancemarketingworld.com/article/1809066/meta-forced-pay-725m-cambridge-analytica-data-scandal},
  Accessed: 2025-03-24, 2018.

\bibitem{owlesg2024}
{OWL AI}, ``{AI's Environmental Impact: Balancing Innovation with
  Sustainability}.''
  {\url{https://owlesg.com/2024/03/26/ais-environmental-impact-balancing-innovation-with-sustainability/},
  Accessed: 2025-03-25}, 2024.

\bibitem{maciejewska2024}
{Joanna Maciejewska}, ``{You know what the biggest problem with pushing
  all-things-AI is? Wrong direction}.''
  \url{{https://x.com/AuthorJMac/status/1773679197631701238}}, Tweet posted on
  March 29, 2024, 2024.

\bibitem{NYTimes2024}
N.~Y. Times, \emph{{Tech Companies' AI Spending Surges Amid Competition}}, .

\end{thebibliography}\endgroup

\end{document}